# Electrically driven magnetism on a Pd thin film

Y. Sun, J. D. Burton and E. Y. Tsymbal

*Department of Physics and Astronomy, Nebraska Center for Materials and Nanoscience,
University of Nebraska, Lincoln, Nebraska 68588-0111, USA*

Using first-principles density functional calculations we demonstrate that ferromagnetism can be induced and modulated on an otherwise paramagnetic Pd metal thin-film surface through application of an external electric field. As free charges are either accumulated or depleted at the Pd surface to screen the applied electric field there is a corresponding change in the surface density of states. This change can be made sufficient for the Fermi-level density of states to satisfy the Stoner criterion, driving a transition locally at the surface from a paramagnetic state to an itinerant ferromagnetic state above a critical applied electric field, $E_c$. Furthermore, due to the second-order nature of this transition, the surface magnetization of the ferromagnetic state just above the transition exhibits a substantial dependence on electric field, as the result of an enhanced magnetoelectric susceptibility. Using a linearized Stoner model we explain the occurrence of the itinerant ferromagnetism and demonstrate that the magnetic moment on the Pd surface follows a square-root variation with electric field, $m \propto (E - E_c)^{1/2}$, consistent with our first-principles calculations.

## I. INTRODUCTION

Research into the coupling between the electric and magnetic order parameters has been highly focused in recent years due to potential applications in high density magnetic recording and spintronic devices.[1,2] This magnetoelectric (ME) coupling is generally characterized by an induction of magnetization by an electric field or, vice versa, induction of an electric polarization by a magnetic field.[3] ME phenomena can be further generalized to electrically controlled exchange bias,[4,5] magnetocrystalline anisotropy[6-8] and the effect of ferroelectricity on spin transport.[9-13]

There are several mechanisms giving rise to ME effects (for recent reviews see Ref [14] and Ref [15]). For example a ME coupling may be observed in single phase compounds if time reversal and space-inversion symmetries are absent.[16,17] However, materials exhibiting this kind of ME coupling are rare and have a comparatively small effect.[18] Currently there is a search for more promising materials systems. For example, a design of novel multiferroic materials with a strong ME coupling was proposed by exploiting epitaxial strain and spin-phonon coupling.[19] A stronger ME effect may also occur in multiphase composites of electrostrictive (or piezoelectric) and magnetostrictive (or piezomagnetic) compounds, mediated through strain across interfaces.[20] In such composite materials, strain induced in one component due an applied electric field is transferred to the other component where, due to the magnetostrictive response, there is a corresponding change in the magnetization.[6,7,21]

Recently researchers have been exploring other routes to ME coupling that may occur at surfaces or interfaces due to pure electronic origins. One of the suggested mechanisms for this kind of ME effect follows from the first-principles study of the Fe/BaTiO$_3$ interface,[22] which shows that the change in bonding between atoms across the interface leads to a change in interface magnetization when the polarization of the ferroelectric BaTiO$_3$ is reversed.[22,23] The ME effect due to the interface bonding mechanism is also expected to play a role for Co$_2$MnSi/BaTiO$_3$ [24] and Fe$_3$O$_4$/BaTiO$_3$ [25] interfaces. Another suggested route is due to the build-up of spin-polarized free-carriers near the surface or interface of a ferromagnetic metal.[26] In such a case, a net change in spin density can be found on the ferromagnetic metal surface/interface due to the spin-dependent nature of the screening charges.[26-29] In some systems more exotic effects due to the accumulation of screening charge may also give rise to substantial ME effects. For example, it was predicted that the change in screening charge at the La$_{1-x}$Sr$_x$MnO$_3$/BaTiO$_3$ interface due to reversal of the ferroelectric polarization can drive a magnetic reconstruction near the interface, changing the magnetic order of Mn spins from ferromagnetic to antiferromagnetic.[30] It was also predicted that even the magnitude of exchange splitting giving rise to itinerant ferromagnetism can be modulated by electrostatic screening. This effect was demonstrated from first-principles at the SrRuO$_3$/BaTiO$_3$ interface,[31] where SrRuO$_3$, a relatively weak itinerant ferromagnet, experiences a pronounced change in exchange splitting upon reversal of the ferroelectric polarization in the BaTiO$_3$. This prediction has led us to explore the possibility of the electric-field-induced itinerant ferromagnetism at an otherwise paramagnetic metal surface.

The feasibility of this ME phenomenon follows from the Stoner criterion for itinerant magnetism. The Stoner model predicts that ferromagnetism in a metal originates from a gain in the total electronic energy, which the sum of the band (kinetic) energy and the exchange energy, with exchange splitting of the spin bands and developing a spontaneous magnetization.[32] The transition from a paramagnetic (PM) to a ferromagnetic (FM) state is determined by the Stoner criterion,

$$I\rho_F > 1 \qquad (1)$$

where $I$ is the Stoner parameter (the exchange constant), and $\rho_F$ is the density of states (DOS) at the Fermi energy. For some transition metals and their alloys the Stoner criterion is close to being satisfied so that the value of $I\rho_F$ is very close to unity. For these metals, an external electric field may induce a surface charge sufficient to lift the DOS at the Fermi energy and, according to the Stoner criterion for magnetism, induce a PM to FM transition at the surface.[33] For example, it was proposed that by alloying FM and PM 3d metals in the appropriate proportion one can create a PM alloy close to ferromagnetic phase transition so that applying an electric field may induce ferromagnetism on the surface of this metal.[34]

Palladium, Pd, is a very good candidate for this kind of the ME effect due to a very large paramagnetic DOS at the Fermi energy.[35,36] In recent years, a lot of efforts have been invested to explore the magnetism in Pd. Both the experiments and theory suggest that magnetism of Pd is very sensitive to the local atomic structure and surrounding environment.[37-45] First-principles studies predicted the magnetism in bulk Pd when increasing its lattice constant by about 4%–6% due to the localization of the 4d-orbital DOS with increasing the separation between atoms[38] and in thin-film Pd of particular thickness due a quantum confinement effect.[45] Magnetism in Pd was predicted and/or observed also in ultrathin films, nanowires, clusters and crystal structures different from fcc.[38-44]

In this paper, we employ first-principles calculations to explore the ME effect on an f.c.c. Pd (001) surface. By explicitly introducing an electric field in our density-functional calculations we demonstrate that the Pd surface can exhibit a PM-FM phase transition due to electrostatically induced screening charge. This extends previous studies of the ME effect on the Pd surface (interface) to a realistic fully self-consistent calculation of the Pd thin film on an appropriate substrate in the presence of an electric field. Using a simple model based on the Stoner approach to itinerant magnetism we explain the results of our first-principles calculations and elucidate the nature of this ME effect.

## II. STRUCTURES AND METHODS

To explore the ME effect of the Pd surface we consider an atomic structure which consists of a thin f.c.c. Pd (001) layer deposited on an f.c.c. Ag (001) substrate, as shown in Fig. 1. This choice is motivated by the previous theoretical work which demonstrates that the magnetism in Pd may be efficiently modulated by Pd layer thickness on a Ag substrate.[46] Therefore, this geometry allows us to choose an appropriate thickness of the Pd layer that produces a surface DOS at the Fermi energy close to satisfy the Stoner criterion (1). In addition, this geometry makes it possible to eliminate the spurious effect of electric field on the bottom Pd surface which would be the case if we used a free standing slab of Pd in our supercell calculation. We find that this condition is satisfied for two Pd monolayers deposited on 5 monolayers of Ag and therefore we keep this geometry in all our calculations (see Fig. 1).

We perform first-principles calculations using the Plane-Wave Self-Consistent Field (PWscf) code based on density functional theory (DFT),[47,48] as implemented in the Quantum-ESPRESSO package.[49] We treat the exchange and correlation by using the local density approximation (LDA). A cut-off energy of 800 eV is used in the calculation and a 34×34×1 **k**-point mesh in the Brillouin zone is necessary for the convergence of the total energy and magnetic moment. The electric field is imposed in the system by using a sawtooth potential as part of the PWscf code.

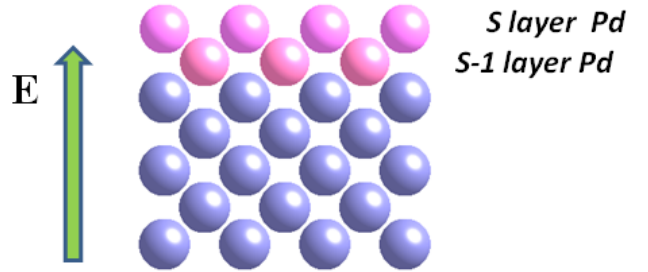

FIG. 1. Atomic structure of a Pd (001) bilayer deposited on 5 monolayers of the fcc Ag (001) substrate, which is used in calculations. The applied electric field (denoted by the arrow) points away from the Pd surface to deplete electrons from the Pd atoms.

Bulk Pd and Ag metals have the experimental lattice constants of 3.89 Å and 4.09 Å, respectively. Within the (001) plane, we constrain the in-plane lattice constant to that of bulk Ag, i.e. 4.09 Å. This expands the Pd lattice constant by about 5% in the (001) plane. Under this constraint, the superlattice is relaxed along the epitaxial [001] direction of Pd growth, until a maximum force on each atom in this direction is less than 0.02 eV/Å. The distance along the [001] direction between the Pd/Ag slabs is kept about 20 Å, which is about half of the superlattice constant. This thickness of the vacuum layer is sufficient to eliminate the unwanted coupling between top and bottom surfaces of the slab and introduce an electric field in our calculations.

## III. RESULTS

Fig. 2 shows the local densities of states (LDOS) projected on the surface (S) and subsurface (S – 1) monolayers of Pd atoms. The calculation reveals that the system is non-magnetic in zero electric field. This result is consistent with the Stoner criterion for magnetism (1). The Stoner parameter $I$ can be estimated from the constrained moment calculation. The latter is performed by evaluating the magnetic (exchange) energy $-¼Im^2$ as the difference between the total energy given by DFT calculation and the band energy for bulk f.c.c. Pd with a constrained moment $m$ per atom.[50] The estimated value is $I \approx 0.65$ eV. As a result, $I\rho_F \approx 0.96$ for the surface Pd atoms which, according to the Stoner criterion (1), makes the surface non-magnetic.

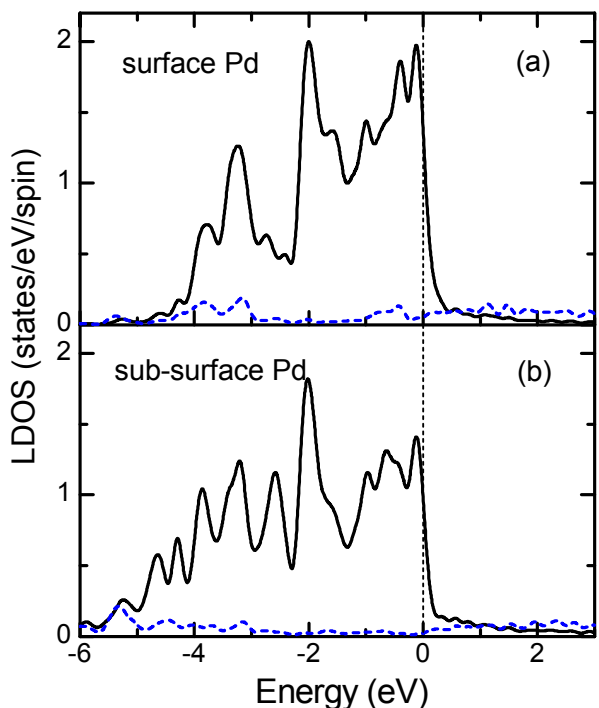

FIG. 2. Local densities of states (LDOS) projected on the surface (a) and sub-surface (b) Pd atoms in the Pd/Ag(001) slab in zero applied electric field, where the system is non-magnetic. Solid curves are the $d$ states and dotted curves are the $s$ states. The vertical dashed line denotes the Fermi energy.

From this we expect that a ~4% enhanced $\rho_F$ should drive the system to ferromagnetism. As follows from Fig. 2, a peak in the Pd surface DOS lies slightly below the Fermi energy. This implies that the enhancement of $\rho_F$ can be realized by a depletion of bands corresponding to lowering the Fermi energy. This can be achieved by applying an external electric field oriented outward the Pd surface to produce a positive screening charge on that surface. Pd is a good metal and therefore it is expected that the screening should occur within roughly one atomic monolayer.

We perform self-consistent DFT calculations in the presence of applied electric field as described in Section II. Fig. 3a demonstrates the shape of electrostatic potential used to generate the applied electric field. Fig. 3b shows the profiles of the induced screening charge density averaged over the cross sectional area of the supercell and calculated for different values of electric field. It is seen that the amplitude of the screening charge on the Pd surface increases with the field. We also note the presence of the Friedel-like oscillation of the induced change density decay into the slab, which is typical for surface electronic screening.[51] A double peak for the screening charge on the Ag surface is also seen in Fig. 3b when the field is above 1.2V/Å. This is due to the potential well lying close to the silver surface that tends to attract the electrons to the dip of the potential energy located at $z = 29$Å. This charge accumulation in vacuum is an artifact of the calculation which is performed using periodic boundary conditions and hence requires a sawtooth potential to introduce an electric field. This does not, however, affect the screening charge on the Pd surface. The latter fact is evident from the comparison of the total screening charge, $\sigma_{calc}$, as a function of applied electric field with the screening charge, $\sigma = \varepsilon_0 E$, expected from the elementary electrostatic theory. The screening charge, $\sigma_{calc}$ is obtained by integrating the planar averaged charge density along the $z$ axis from the vacuum half way through to the middle of the slab. The results are displayed in Fig. 4 and show good consistency. The small deviation of $\sigma_{calc}$ from $\varepsilon_0 E$ is the consequence of an intrinsic charge transfer between the two metals in contact that establishes a constant chemical potential across the Pd/Ag system. As follows from the calculation, the magnitude of the contact charge is about $3\times10^{-4}$ $e$/Å$^2$.

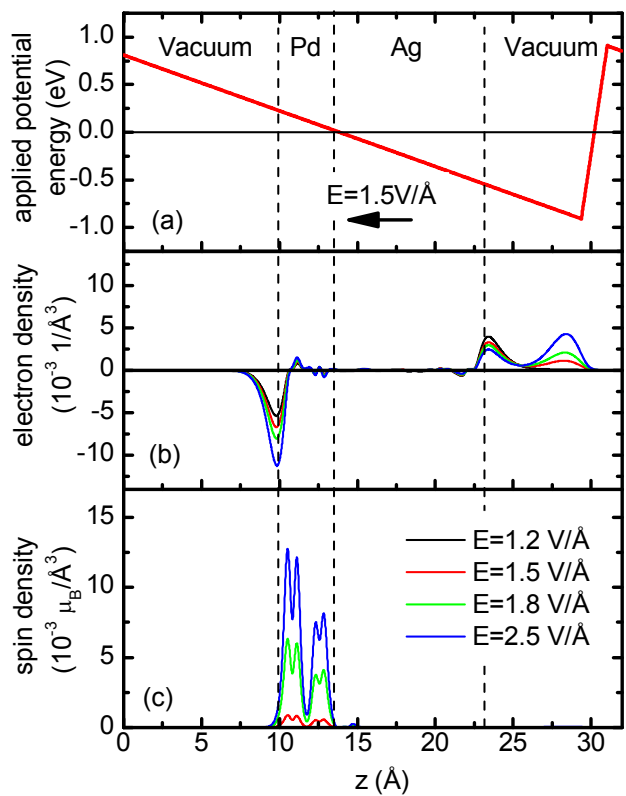

FIG. 3. (a) The electrostatic potential energy $-e\phi(z)$ used in the calculations to generate an applied electric field $E$ of 1.5 V/Å. (b) The induced electron densities, $\Delta n = n(E) - n(0)$, averaged over the $x$-$y$ plane and plotted along the $z$ direction normal to the surface. The dashed lines are used to show schematically regions of vacuum, Pd and Ag layers. (c) The planar averaged spin density along the $z$ axis.

As predicted, the application of the sufficiently large electric field induces the magnetic moment of the Pd atoms. This is evident from Fig. 3c which shows the planar averaged spin densities for different fields resolved on the $z$ axis. It is seen that the spin density is zero for $E = 1.2$ V/Å, whereas for 1.5 V/Å and above there is an induced spin-density on the Pd atoms. The complete dependence of Pd magnetic moments on electric field is displayed in Fig. 5. At a critical field of $E_c = 1.5$ V/Å a PM to FM transition takes place on the Pd surface according to the Stoner criterion, as discussed in detail in the next section.

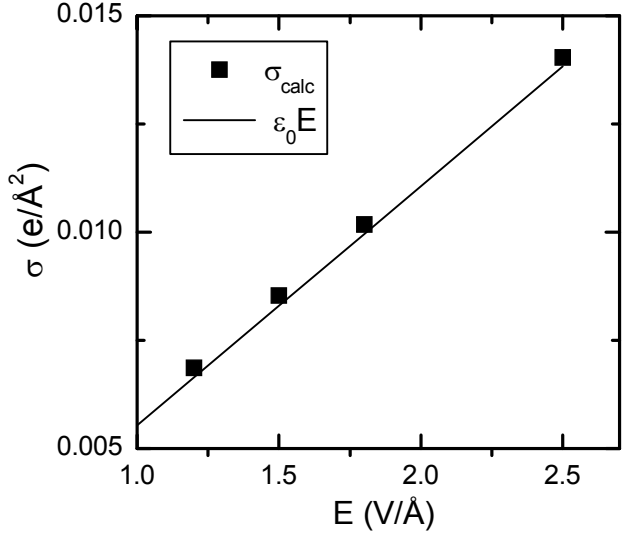

FIG. 4. The screening charge densities calculated by integration of the planar averaged surface charge density, $\sigma_{calc}$ (squares), and obtained from the electrostatic theory, $\sigma = \varepsilon_0 E$ (line).

## IV. ORIGINS OF ELECTRICALLY INDUCED MAGNETISM

We apply the results of a linearized Stoner model (derived in the Appendix) to the surface Pd atom given the fact that the screening charge is largely localized within the surface Pd monolayer. To produce a magnetic moment on a paramagnetic Pd surface the electric field needs to be applied outward from the Pd surface to deplete electrons from the surface and hence enhance the LDOS (see Fig. 2). This depletion may be thought as a shift of the paramagnetic LDOS with respect to the Fermi energy. This shift can be estimated as

$$\delta\varepsilon \approx -\frac{\varepsilon_0 E A}{2e\rho_F}, \quad (2)$$

where $A$ is the cross sectional area of the supercell and $\rho_{F0}$ is the paramagnetic DOS (per spin) at the Fermi energy in the absence of an applied electric field. Since the shift is small we can assume that the corresponding change of the LDOS at the Fermi level is much less than the LDOS itself, i.e., $\delta\rho_F \ll \rho_F$, so that

$$\rho_F(E) = \rho_F + \rho'_F \delta\varepsilon = \rho_F + \alpha E, \quad (3)$$

where $\rho'_F$ is the first derivative of the paramagnetic DOS at the Fermi level, which we assume to be independent of the electric field, and

$$\alpha = -\frac{\rho'_F}{2e\rho_F}\varepsilon_0 A. \quad (4)$$

Now we assume that there is a critical field $E_c$ that triggers the magnetism in the system, so that at this field the Stoner criterion becomes fulfilled, i.e.

$$(\rho_F + \alpha E_c)I = 1. \quad (5)$$

Using Eq. (3) in Eq. (A13) for the magnetic moment in the linearized Stoner model, and taking $\delta\rho_F \ll \rho_F(E)$ into account, we obtain the expression for the magnetic moment as a function of the electric field,

$$m(E) = \frac{2}{I|\rho'_F|}\sqrt{2\alpha(E - E_c)\rho_F}. \quad (6)$$

This is the main result of the linearized Stoner model which we now compare with our first-principles calculations.

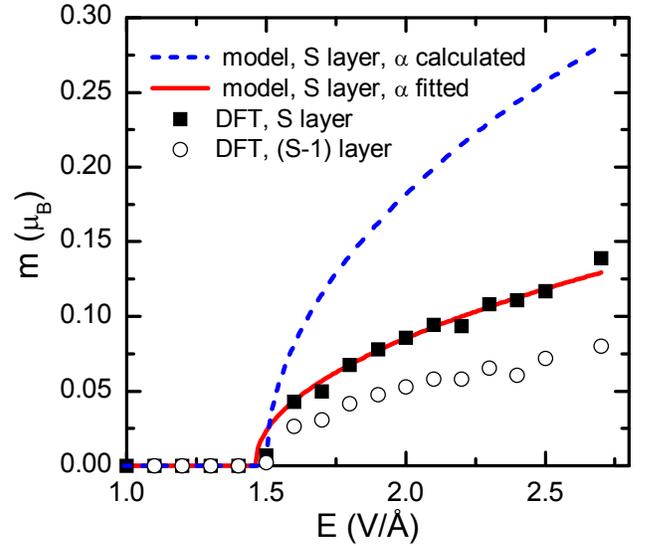

FIG. 5. Magnetic moment as a function of electric field. Symbols display the magnetic moment from the first-principles calculation for the surface (S) (squares) and the sub-surface $(S-1)$ (circles) Pd monolayer. Lines display results of the theoretical model (6) where the parameter $\alpha$ is calculated from Eq. (4) using $I = 0.61$ eV (dashed line) or obtained from the slope of the linear fit in Fig. (6) using $I = 0.66$ eV (solid line).

In Fig. 6 we show by black squares the induced magnetic moment on the surface Pd atom calculated from first principles as a function of applied electric field. A pronounced offset of magnetism is seen at $E_c = 1.5$ V/Å. The dashed line shows the result of our model (14) where the parameter $\alpha = 0.12$ Å/eV$^2$ is obtained from Eq. (4) using parameters from the band calculation. At the same time, the critical field in Eq. (5) is adjusted to 1.5 V/Å by calling the Stoner parameter to be 0.61eV, which is reasonably close to the estimated value, 0.65eV, from our constrained moment calculation. As is seen from Fig. 5, the plots from the model (dashed line) and first-principles calculations (black squares) display a similar tendency, however, quantitatively the model overestimates the first-principles result by a factor of two.

This disagreement can be understood in view of results displayed in Fig. 6, which demonstrates the variation of the local DOS at the Fermi energy as a function of electric field. The assumption of the local DOS at the surface Pd monolayer being entirely determined by the screening charge and the associated shift in the paramagnetic DOS leads to the strongly overestimated change in the DOS with increasing the field (dashed line in Fig. 6) as compared to the first-principles calculation (squares in Fig. 6). This is due to hybridization between the $d$ orbitals of the surface and sub-surface Pd atoms which is evident from the induced change of the sub-surface Pd LDOS with electric field (see circles in Fig. 6).

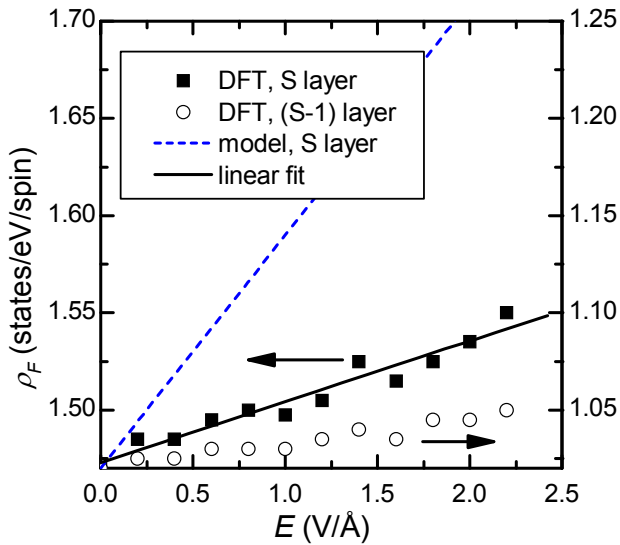

FIG. 6. Density of states at the Fermi energy for paramagnetic Pd as a function of electric field. Symbols display the DOS from the first-principles calculation for the surface (S) (squares) and the sub-surface (S – 1) (circles) Pd monolayer. The dashed line displays the result of Eqs. (3) and (4). The solid line is the linear fit to the DFT calculation.

Therefore, in order to verify the correctness of the linearized Stoner model we have to use the LDOS consistent with of our DFT calculation. For this purpose, we fit the DOS on the surface Pd monolayer using a linear function, as shown in Fig. 6 by the solid line. Using the value of $\alpha$ obtained from this fit and Eq. (5) we can fully reproduce the variation of the induced magnetic moment as a function of electric field using $I = 0.66$eV (see the solid line in Fig. 5). This value is consistent with the value of $I = 0.65$eV obtained from the fixed moment calculation in the bulk.

In addition to the surface Pd atom, we find the electric field induced magnetism also on the sub-surface Pd atom. From Fig. 6, we see that the sub-surface palladium atom has a much lower DOS at the Fermi level than the surface atom. The sub-surface palladium atom never meets the Stoner criterion ($I\rho_F \approx 0.7$), however, as is evident from Fig. 3c and Fig. 5 (circles), there is a sizable magnetic moment induced on this atom above the critical electric field. We see also that the DOS at the Fermi level on the sub-surface monolayer changes with the electric field, even though the screening charges are almost entirely localized on the surface monolayer (see Fig. 3b). The change in the electronic DOS induced by the screening charge on the surface Pd monolayer is partly transferred to the subsurface Pd monolayer as the result of hybridization between the two neighboring Pd atoms. This effect leads in the induced magnetization on the sub-surface Pd monolayer similar to that known from the previous calculations on different systems (see, e.g., ref. [52] where a magnetic Co atom is predicted to induce a magnetic moment of the adjacent Ti atom).

The predicted variation of the magnetic moment as a function of applied electric field is distinctly different from the previous predictions of electrically induced magnetism of electronic origin. On magnetic surfaces or interfaces with dielectric materials the induced magnetization is linear with electric field.[26-29] At the ferromagnet/ferroelectric interfaces the variation is non-linear and follows the hysteretic behavior of the spontaneous polarization in the ferroelectric.[22-25,30,31] In the case of a paramagnetic surface close to the Stoner criterion the electric field drives the transition locally at the surface from a paramagnetic state to an itinerant ferromagnetic state above a critical applied electric field. This variation of the magnetic moment with the field follows Eq. (6), which exhibits a second-order transition, where the surface magnetization of the ferromagnetic state just above the transition exhibits a substantial dependence on electric field, resulting in a divergence of the magnetoelectric susceptibility at the transition.

In order to observe the predicted phenomenon very large electric fields are required which may be achieved using scanning probe techniques such as STM or conducting tip AFM. Within the latter it may be efficient to use a high $\kappa$ dielectric on the paramagnetic metal surface to enhance the screening charge due to the large dielectric permittivity. Another route is to investigate paramagnet/ferroelectric interfaces where the screening charge is expected to be dramatically enhanced due to the large polarization charge in the ferroelectric. Exploring theoretically paramagnet/dielectric(or ferroelectric) interfaces in this regards would be interesting.

## V. SUMMARY

We have demonstrated the principal feasibility to produce a paramagnetic-ferromagnetic (PM-FM) phase transition on the surface of a paramagnetic metal by an applied electric field. We performed self-consistent density-functional electronic structure calculations in the presence of external electric field using Pd/Ag(001) system as a representative example. We found an induced magnetic moment on the surface Pd monolayer when the electric field exceeds a critical field. The effect occurs due to the induced screening charge at the surface which enhances the local density of states at the surface Pd monolayer and, in accordance with the Stoner criterion for magnetism, drives the PM-FM transition. Using a linearized Stoner model we explained the occurrence of the ferromagnetism and demonstrated that the magnetic moment follows a square-root variation with electric field consistent with our first-principles calculations. This predicted PM-FM transition manifests the second order phase transition at which a substantial dependence on electric field is expected as the result of an enhanced magnetoelectric susceptibility.

## APPENDIX: LINEARIZED STONER MODEL

The basic idea of the Stoner model is as follows. Due to the localized nature of $d$ states, two $d$ electrons will experience a strong Coulomb repulsion if they occupy the same orbital, in which case they must have opposite spin orientations due to the Pauli exclusion principle. To reduce this Coulomb energy it is advantageous for the $d$ electrons to instead have parallel spins and to occupy different orbitals. To accomplish this, however, one of the electrons must be transferred to a previously unoccupied orbital with higher kinetic energy. The Stoner model encompasses this competition between Coulomb repulsion, the Pauli exclusion principle and increased kinetic energy explicitly for continuous bands of electronic states.

If there is an exchange splitting $\Delta$ of the spin bands, there is a formation of a spontaneous magnetic moment:

$$m = \int_{\varepsilon_F - \Delta_2}^{\varepsilon_F + \Delta_1} \rho(\varepsilon) d\varepsilon \ . \quad (A1)$$

Here $\varepsilon_F$ is the Fermi energy, $\rho(\varepsilon)$ is the DOS per spin in the paramagnetic state, which is assumed to be rigid. $\Delta_1$ and $\Delta_2$ denote the exchange driven shifts of the majority and minority-spin bands with respect to the spin bands in the paramagnetic state so that

$$\Delta_1 + \Delta_2 = \Delta \ . \quad (A2)$$

The gain in the exchange energy enters through a phenomenological term given by

$$U_{ex} = -\tfrac{1}{4} I m^2 \ , \quad (A3)$$

where $I$ is the Stoner parameter (exchange constant) characterizing the strength of the Coulomb repulsion.

However, moving electrons from occupied states of one spin channel to unoccupied states of the opposite spin direction necessarily enhances the total kinetic energy,

$$U_{kin} = \int_{\varepsilon_F - \Delta_2}^{\varepsilon_F + \Delta_1} \varepsilon \rho(\varepsilon) d\varepsilon \ . \quad (A4)$$

There are, therefore, two competing tendencies, which have to be balanced in order to find whether ferromagnetism is favored. The stability condition which has to be satisfied for the appearance of ferromagnetism is the Stoner criterion (1). The Stoner criterion, however, only indicates whether or not the paramagnetic state is stable with respect to the formation of an exchange split ferromagnetic case. To actually determine the equilibrium exchange splitting $\Delta$ and magnetic moment $m$ arising from this competition between Coulomb repulsion and kinetic energy one must minimize the total energy $U = U_{ex} + U_{kin}$, which leads to the well-known relation for the stabilized ferromagnetism[50]

$$m = \frac{\Delta}{I} \ . \quad (A5)$$

Of course this deceptively simple relation depends implicitly on the details of the paramagnetic DOS which, in general, may have a complicated dependence on energy, $\varepsilon$.

In our case, when the exchange splitting is small and all the changes occur in the vicinity of the Fermi energy, we can linearize the DOS in this region of energies so that

$$\rho(\varepsilon) = \rho_F + (\varepsilon - \varepsilon_F) \rho'_F \ , \quad (A6)$$

where $\rho_F$ and $\rho'_F$ are the paramagnetic DOS per spin and its derivative at the Fermi energy. Fig. 7 illustrates schematically the linearization scheme. Due to the charge conservation we have

$$\int_{\varepsilon_F}^{\varepsilon_F + \Delta_1} \rho(\varepsilon) d\varepsilon = \int_{\varepsilon_F - \Delta_2}^{\varepsilon_F} \rho(\varepsilon) d\varepsilon \ , \quad (A7)$$

which within the linear approximation (A6) leads to

$$\rho_F \Delta_1 + \tfrac{1}{2} \Delta_1^2 \rho'_F = \rho_F \Delta_2 - \tfrac{1}{2} \Delta_2^2 \rho'_F \ . \quad (A8)$$

Using eq. (A2) we find

$$\Delta_{1,2} = \frac{\Delta}{2} \mp \left( \frac{\rho_F}{\rho'_F} - \sqrt{\left(\frac{\rho_F}{\rho'_F}\right)^2 - \frac{\Delta^2}{4}} \right) \ . \quad (A9)$$

The linear approximation (A6) results in the magnetic moment (A1)

$$m = \rho_F \Delta + \tfrac{1}{2}(\Delta_1^2 - \Delta_2^2) \rho'_F \ , \quad (A10)$$

which modifies the relation (A5) as follows

$$I \rho_F + \tfrac{1}{2}(\Delta_1 - \Delta_2) I \rho'_F = 1 \ . \quad (A11)$$

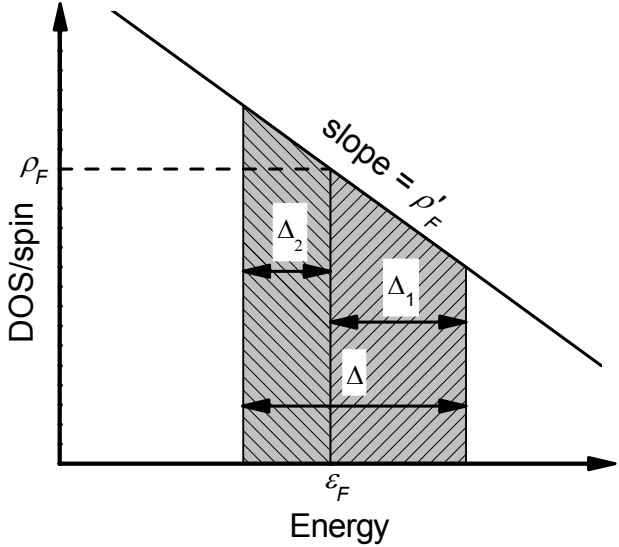

FIG. 7. Schematic of the linearized Stoner model. In a narrow range around the Fermi level, $\varepsilon_F$, we treat the paramagnetic DOS (per-spin) as a linear function (solid line) with slope $\rho'_F$ passing through $\rho_F$ at $\varepsilon_F$. An exchange splitting, $\Delta$, is introduced such that a number of minority-spin electrons with energies down to $\Delta_2$ below $\varepsilon_F$ are transferred to the majority-spin band up to an energy $\Delta_1$ above $\varepsilon_F$. $\Delta_1$ and $\Delta_2$ are mutually constrained by the charge conservation condition, i.e. the areas of the two shaded regions above and below $\varepsilon_F$ are equal.

Using Eq. (A9) in (A11) we can now solve for the equilibrium exchange splitting

$$\Delta = \frac{2}{I|\rho'_F|}\sqrt{I^2\rho_F^2 - 1} \ . \quad (A12)$$

The absolute value sign emerges here because it is convenient to assume that the value of $\Delta$ is positive. This allows us to express the magnetic moment in terms of $\rho_F$

$$m = \frac{2}{I^2|\rho'_F|}\sqrt{I^2\rho_F^2 - 1} \ . \quad (A13)$$

This expression holds for small magnetic moments and exchange splitting where the paramagnetic DOS can be approximated to be linear near the Fermi energy. Of course in the case we are interested in $\rho_F$ depends implicitly on the magnitude of an applied electric field $E$, and indeed this is the origin of the effect discussed in this paper.

We note that the Stoner model is rigorously valid only for a homogeneous system. For inhomogeneous systems a more sophisticated approach based on magnetic susceptibility is required.[50] As seen in Section IV, however, our simple treatment is consistent with our first-principles calculations providing a transparent interpretation of the predicted phenomenon.

### ACKNOWLEDGMENTS

This work was supported by the NSF-MRSEC (grant No. DMR-0820521), the Nanoelectronics Research Initiative of the Semiconductor Research Corporation and the Nebraska Research Initiative. Computations were performed utilizing the Research Computing Facility at UNL and the Center for Nanophase Materials Sciences at Oak Ridge National Laboratory.